\documentclass[conference]{IEEEtran}
\IEEEoverridecommandlockouts
\usepackage{cite}
\usepackage{amsmath,amssymb,amsfonts}
\usepackage{algorithmic}
\usepackage{graphicx}
\usepackage{textcomp}
\usepackage{xcolor}
\def\BibTeX{{\rm B\kern-.05em{\sc i\kern-.025em b}\kern-.08em
    T\kern-.1667em\lower.7ex\hbox{E}\kern-.125emX}}
\begin{document}

\title{Audio-Based Deep Learning Frameworks for Detecting COVID-19
}
\author{Dat~Ngo$^{1}$, 
             Lam~Pham$^{2}$, 
             Truong~Hoang$^{3}$,
             \c{S}efki~Kolozali$^{1}$,
             Delaram~Jarchi$^{1}$ \\ \\
1. School of Computer Science and Electronic Engineering, University of Essex, United Kingdom.\\
2. Center for Digital Safety \& Security, Austrian Institute of Technology, Austria.\\

3. Solution Technology Unit Department, FPT Software Company Limited, Vietnam. \\

}

\maketitle

\begin{abstract}
This paper evaluates a wide range of audio-based deep learning frameworks applied to the breathing, cough, and speech sounds for detecting COVID-19. In general, the audio recording inputs are transformed into low-level spectrogram features, then they are fed into pre-trained deep learning models to extract high-level embedding features. Next, the dimension of these high-level embedding features are reduced before fine-tuning using Light Gradient Boosting Machine (LightGBM) as a back-end classification. 
Our experiments on the Second DiCOVA Challenge achieved the highest Area Under the Curve (AUC), F1 score, sensitivity score, and specificity score of 89.03\%, 64.41\%, 63.33\%, and 95.13\%, respectively. Based on these scores, our method outperforms the state-of-the-art systems, and improves the challenge baseline by 4.33\%, 6.00\% and 8.33\% in terms of AUC, F1 score and sensitivity score, respectively.
\end{abstract}


\section{Introduction}

The COVID-19 pandemic is now continuing to spread around the world, with more than 300 million confirmed cases, and causing more than five million deaths across almost 200 countries~\cite{bbc}. 
Although many countries are in relentless efforts of launching their vaccination programs, the number of global daily new cases is still increasing as COVID-19 restrictions are being eased in many countries and people can travel worldwide. 
To deal with the rapid spread of infection across populations, many countries have to conduct massive daily tests for their citizens to control the virus spreads. 
Particularly, the molecular testing approaches, namely the reverse transcription polymerase chain reaction test (RT-PCR)~\cite{corman2020detection} and the rapid antigen test (RAT)~\cite{peeling2021scaling}, are now widely applied as primary testing methodologies in most countries. However, these methodologies present various limitations as their procedure of sample collection violates physical distancing in many countries that the ordering online lateral test at home is limited. Additionally, analysing and receiving results require high-cost, chemical equipment, and involving labour-intensive tasks. 
As a result, there is an urgent need for non-invasive, scalable, and cost-effective tool to detect infected individuals in a decentralized manner.
As COVID-19 disease is related to primary symptoms such as fever, sore throat, cough, chest pain, etc. Therefore, many researchers and practitioners are motivated to use acoustic-related signals such as breathing, cough, speech from human respiratory system to early detect COVID-19 and non-COVID-19 individuals~\cite{andreu2021generic}. 
Indeed, calls for development of diagnostic tools were announced in the Interspeech 2021 as a special session titled `Diagnostics of COVID-19 using Acoustics (DiCOVA) Challenge' as well as in ICASSP 2022 as a calling paper, referred to as the First DiCOVA and the Second DiCOVA Challenges, respectively.

In this paper, we aim to explore all types of human respiratory sounds: breathing, cough, speech, provided by the recent Second DiCOVA Challenge, and then propose a robust framework for detecting COVID-19. 
Our contributions include: (1) to conduct extensive experiments and pinpoint the most effective approach related to the extraction of well-represented features for each type of breathing, cough, speech sound input; (2) to evaluate how oversampling on positive samples and dimension reduction on represented features affect the system performance; and (3) to demonstrate that our proposed framework is reliable and robust for detecting COVID-19 which outperforms the state-of-the-art systems and is potential for a real-life application.

\section{Dataset and tasks defined}
\label{dataset}

In this paper, we evaluate the dataset derived from the Second DiCOVA Challenge~\cite{muguli2021dicova}. 
This dataset provides audio recordings of three different sound categories: breathing, cough, speech which were collected from both COVID-19 positive and negative patients in the age group of 15 to 45 years old. 
For each sound category, there are a total of 1436 acoustic samples which are then divided into the Development set (965 subjects) and the Test set (471 subjects) for training and testing processes, respectively. It is of note that Development set in each categories has recordings of 172 individuals positive to COVID-19, while there is a higher number of recordings from 793 individual negative to COVID-19, presenting an imbalanced dataset. 
To compare our proposed systems with the state-of-the-art, we follow the Second DiCOVA Challenge.
Particularly, we aim to detect COVID-19 positive subjects by exploring only breathing sound, only cough sound, only speech sound, and using all sound categories, which match the Track-1, Track-2, Track-3, and Track-4, respectively in the Second DiCOVA Challenge. 
Regarding the evaluation metrics, we also obey the Second DiCOVA Challenge which uses An Area Under the Curve (AUC) and the specificity (SPEC.) \& sensitivity (SEN.).
As SPEC. has to be equal or larger than 95.13\% which is an essential requirement of the challenge, a decision threshold from 0 to 1 with a step size of 0.0001 is evaluated to obtain SEN. when SPEC. value is satisfied.  
As a result, we report AUC and SEN. in this paper (i.e. The SPEC. is always equal or greater than 95.13\% to meet the challenge's requirement as presented in the leaderboard of the challenge~\cite{leaderboard}). 

\section{The Proposed Framework}
\label{proposedframework}
\begin{figure}[t]
	\centering
	\centerline{\includegraphics[width=1\linewidth]{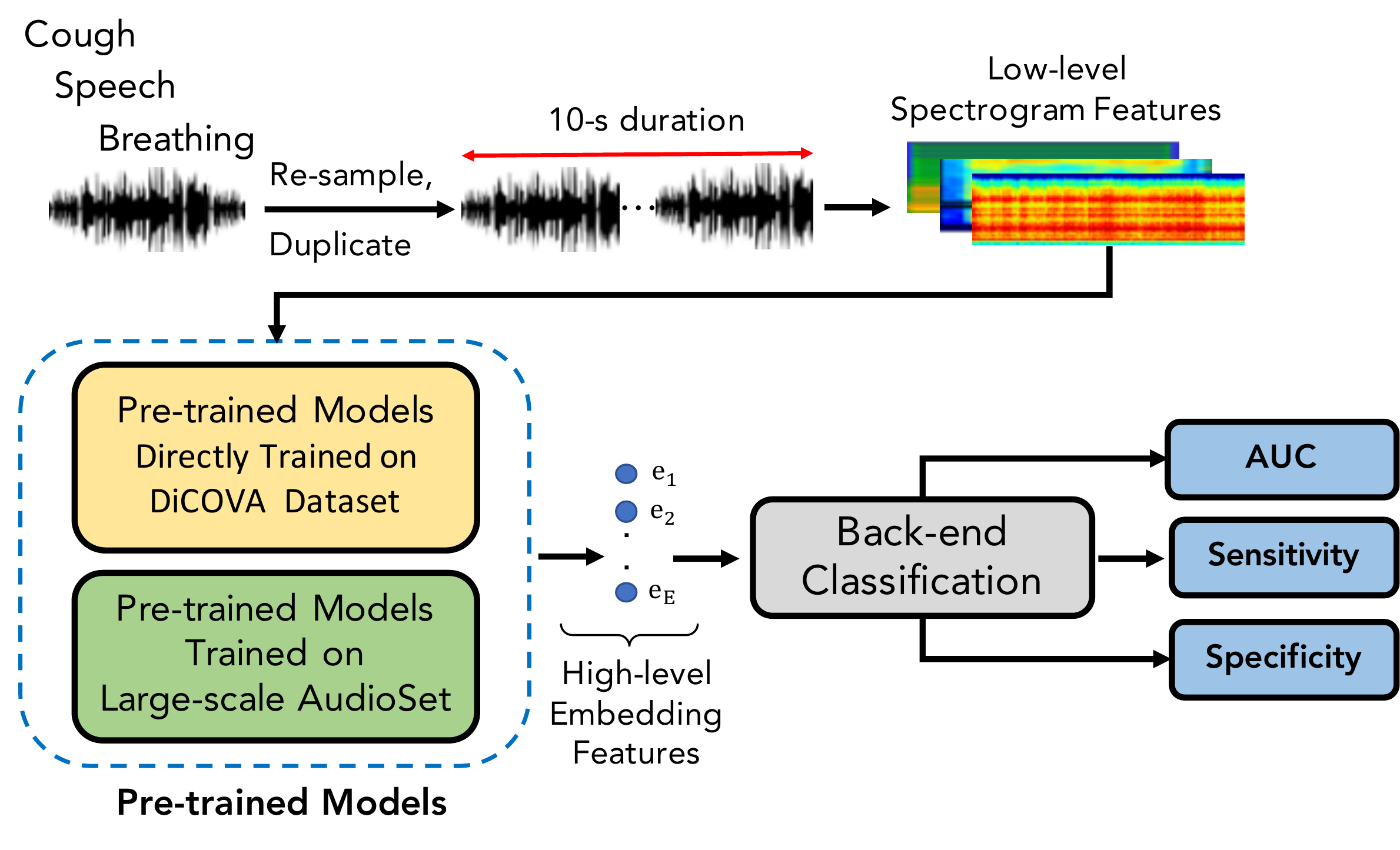}}
	\caption{The high-level architecture of the proposed frameworks}
		 \vspace{-0.3 cm}
	\label{fig:architecture}
\end{figure}	

To explore DiCOVA dataset, we firstly present a high-level architecture of proposed framework as shown in Figure~\ref{fig:architecture}.
Generally, an entire framework proposed is separated into three main steps: Low-level spectrogram feature extraction, high-level embedding feature extraction, and back-end classification.

\subsection{The low-level spectrogram feature extraction}
At the first step shown in the upper stream in Figure~\ref{fig:architecture}, the raw audio recordings are re-sampled to 16.000 Hz, duplicated to make sure that all audio recordings have an equal duration of 10 seconds.
These audio recordings are then transformed into spectrograms where both temporal and spectral features are presented.

\subsection{The high-level embedding feature extraction}
\label{2framework}
The low-level spectrogram features are then fed into pre-trained deep learning models to extract embeddings (e.g. vectors), referred to as high-level features.
The pre-trained deep learning models used for extracting high-level embedding features come from two different approaches: \textbf{(I)} pre-trained models directly trained on DiCOVA dataset and \textbf{(II)} pre-trained models trained with the large-scale AudioSet dataset~\cite{gemmeke2017audio}.

\textbf{(I) In the first approach}, we construct deep learning models and directly train these models on DiCOVA dataset mentioned in Section~\ref{dataset} with Cross-entropy loss and  Adam method~\cite{Adam} for optimization.
After the training process, we reuse these models to extract the feature map at the global pooling layer, which is considered as the high-level embedding features mentioned.
As the performance of high-level embedding features depends on the low-level spectrograms and the model architectures in this approach, we therefore base on our previous work~\cite{pham2022ensemble, ngo2021deep, pham2021cnn, pham2021inception, pham2020robust} to evaluate: three types of spectrograms such as log-Mel~\cite{librosa_tool}, Gammatonegram~\cite{aud_tool}, and Scalogram~\cite{lilly2008higher} that are proven effective in representing respiratory-related sounds; and a wide range of benchmark deep learning models from low footprint models such as Lenet based network~\cite{lecun1999object} to high-complex architectures such as Xception~\cite{chollet2017xception}, InceptionV3~\cite{szegedy2016rethinking}, etc.

To identify which spectrogram features are well-represented for breathing, cough, or speech audio inputs, we evaluate three proposed spectrograms using a Lenet based network architecture as shown in Table~\ref{table:lenet}. 
To make sure that the spectrograms that are fed into the Lenet based model have the same size, we use the same setting parameters with window size = 2048, hop size = 1024, and filter number = 128; to generate the same spectrogram of 128$\times$154.
Regarding the Lenet based architecture as shown in Table~\ref{table:lenet}, it includes the convolutional layer (Conv [kernel size]), batch normalization (BN)~\cite{ioffe2015batch}, rectified linear units (ReLU)~\cite{relu}, average pooling (AP), global average pooling (GAP), dropout~\cite{dropout} (Dr(percentage)), fully connected (FC) and Softmax layer. 
Since certain low-level spectrograms are suitable for various types of audio inputs (i.e. breathing, cough, or speech), we therefore evaluate these spectrograms with different benchmark neural network architectures of VGG16~\cite{simonyan2014very}, VGG19~\cite{simonyan2014very}, MobileNetv1~\cite{howard2017mobilenets}, ResNet50~\cite{he2016deep}, Xception~\cite{chollet2017xception}, InceptionV3~\cite{szegedy2016rethinking}, and DenseNet121~\cite{huang2017densely} to achieve the best framework configuration (i.e. which low-level spectrogram and training network architecture).
As evaluated network architectures are reused from the Keras library~\cite{keras_app}, the final fully connected layer of these networks is modified from 1000 (i.e. the number of classes of image object in ImageNet datset) to 2 which matches the number of classes in DiCOVA dataset.  
It is worth noting that we only reuse the network architectures from the Keras library instead of using available weights trained with ImageNet dataset, all trainable parameters of these networks are initialized with mean and variance set to 0 and 0.1, respectively. 

\begin{table}[t]
    \caption{Configuration of Lenet based architecture} 
        	\vspace{-0.2cm}
    \centering
    \scalebox{0.85}{
    \begin{tabular}{|c | c |} 
        \hline 
            \textbf{Lenet architecture layers}   &  \textbf{Output}  \\
        \hline 
         Input layer (spectrogram patch of $128{\times}154{\times}1$)  &        \\
         BN - Conv [$3{\times}3$] @ 32  - ReLU - BN - AP [$2{\times}2$] - Dr ($20\%$)      & $64{\times}77{\times}32$\\
         BN - Conv [$3{\times}3$] @ 64 - ReLU - BN - AP [$2{\times}2$] - Dr ($25\%$)      & $32{\times}38{\times}64$\\
         BN - Conv [$3{\times}3$] @ 128 - ReLU - BN - AP [$2{\times}2$] - Dr ($30\%$)      & $16{\times}19{\times}128$\\
         BN - Conv [$3{\times}3$] @ 256 - ReLU - BN - GAP - Dr ($35\%$)       & $256$\\
         FC - ReLU - Dr ($40\%$) &  $1024$       \\
         FC - ReLU - Dr ($40\%$) &  $1024$       \\
         FC - Softmax & C=2 \\
       \hline 
    \end{tabular}
    }
    \vspace{-0.3cm}
    \label{table:lenet} 
\end{table}

\textbf{(II) In the second approach}, we leverage three available pre-trained models trained with the large-scale AudioSet dataset in advance: PANN~\cite{kong_pretrain}, OpenL3~\cite{cramer2019look}, TRILL~\cite{shor2020towards}.
We evaluate whether these up-stream pre-trained models are beneficial for the down-stream task of detecting COVID-19 positive in DiCOVA dataset.
While PANN~\cite{kong_pretrain} and OpenL3~\cite{cramer2019look} leverage VGGish architecture and cross-entropy loss to train the large-scale Audioset dataset, TRILL~\cite{shor2020towards} is based on Resnet and tripless loss.
Additionally, while PANN~\cite{kong_pretrain} was trained using input signal with 10-second duration, both TRILL~\cite{shor2020towards} and OpenL3~\cite{cramer2019look} analyse short time duration of 1 second.
As using different network architectures, loss functions, as well as analysing different audio durations, these proposed pre-trained models may differently perform on three types of audio inputs (i.e. breathing, cough, and speech) from DiCOVA dataset.

Regarding the high-level embedding features extracted from these pre-trained models, while OpenL3~\cite{cramer2019look} and PANN~\cite{kong_pretrain} extract the feature map at the global pooling layer, TRILL~\cite{shor2020towards} extracts the feature map at the final layer which proves effective for different down-stream tasks mentioned in~\cite{shor2020towards}. 
Notably, since PANN pre-trained model works on 10-second input duration from AudioSet dataset that matches the input duration of our proposed system in Figure~\ref{fig:architecture}, only one embedding feature (e.g. one vector) is extracted from each 10-second audio recordings of the input.
Meanwhile, as OpenL3 and TRILL pre-trained models were trained with 1-second audio segment of AudioSet dataset, therefore, multiple embeddings are obtained when we feed one 10-second DiCOVA audio sample into these two pre-trained models.
Therefore, an average of these multiple embeddings across the time dimension is computed to obtain one embedding feature which represents for each 10-second duration audio input.
Additionally, only low-level log-Mel spectrogram is used in this approach as these three pre-trained models explore this type of spectrogram for training on the large-scale AudioSet dataset. 

As we present two different approaches of using pre-trained models for extracting high-level embedding features, we now refer to two main frameworks as:  (I) three  low-level  spectrograms (log-Mel, Gammatonegram, Scalogram),  pre-trained  models directly trained on DiCOVA dataset, LightGBM back-end classification; and (II) low-level log-Mel spectrogram, pre-trained models trained on the large-scale AudioSet, and LightGBM back-end classification.
\begin{table}[t]
	\caption{Comparison of low-level spectrograms (AUC/Sen.)} 
        	\vspace{-0.2cm}
    \centering
    \scalebox{0.85}{
    \begin{tabular}{|c | c c c  c|} 
        \hline 
           \textbf{Spec}    &\textbf{Cough} &\textbf{Speech} &\textbf{Breathing} &\textbf{All} \\
        \hline 
	    GAM             &70.11/10.00  &73.41/21.67  &73.29/15.00 &77.38/25.00 \\
	    \textbf{log-MEL}               &\textbf{71.39/16.66}  &\textbf{73.15/18.33}  &\textbf{74.31/18.33} &\textbf{80.11/33.33}\\
	    Scalogram          &63.71/8.33  &68.68/10.00  &78.20/20.00 &69.83/18.33 \\
       \hline 
    \end{tabular}
    }
    \vspace{-0.3cm}
    \label{table:spec_bs} 
\end{table}

\subsection{The back-end classification}
In this paper, we use Light Gradient Boosting Machine (LightGBM)~\cite{ke2017lightgbm} as the final back-end classification model to fine-tune high-level embedding features. 
The LightGBM is implemented by using an available toolkit~\cite{ke2017lightgbm} and the parameters are set as: learning\_rate = 0.02, objective = ‘binary’, metric = ‘auc’, subsample = 0.68, colsample\_bytree = 0.28, early\_stopping\_rounds = 1000, num\_iterations = 10000, subsample\_freq = 1.
As the Track-4 in the Second DiCOVA Challenge suggests to use all audio input data (i.e. breathing, cough, speech), high-level embedding features extracted from different types of audio inputs are concatenated before feeding into the back-end LightGBM for classification.

\section{Experiments and results}
\label{experiment}



\subsection{Performance comparison of frameworks (I): three low-level spectrograms, pre-trained models directly trained on DiCOVA dataset, and LightGBM back-end classification}
\label{framework1}
\begin{table}[t]
	\caption{Comparison of pre-trained deep learning network architectures directly trained on DiCOVA dataset (AUC/SEN.)} 
        	\vspace{-0.2cm}
    \centering
    \scalebox{0.85}{
    \begin{tabular}{|c | c c c c |} 
           \hline 
         \textbf{Input}  &\textbf{Cough} &\textbf{Speech} &\textbf{Breathing} &\textbf{All} \\
        \hline 
	    VGG16                 &70.05/10.00 &75.64/10.00 &71.75/16.66 &78.17/30.00 \\
	    VGG19                 &63.07/5.00 &67.81/8.33 &70.28/15.00 &68.24/6.67\\
	    MobileNetv1           &58.92/5.00 &66.86/16.67 &67.76/13.33 &70.73/25.00\\
	    ResNet50              &66.67/10.00 &70.07/21.67 &69.39/18.33 &77.35/23.33\\
	    \textbf{Xception}              &\textbf{71.21/31.66} &\textbf{70.91/28.33} &\textbf{72.75/28.33} &\textbf{80.81/43.33}\\
	    InceptionV3           &61.97/8.33 &74.39/30.00 &67.56/18.33 &78.83/30.00\\
	    DenseNet121           &68.12/23.33 &75.55/20.00 &72.63/18.33 &80.10/31.66\\
       \hline 
    \end{tabular}
    }
        \vspace{-0.3cm}
    \label{table:pre_train_01} 
\end{table}
We firstly evaluate how low-level spectrograms affect the performance in frameworks (I).
As the results are shown in Table~\ref{table:spec_bs}, log-Mel performs better than Gammatonegram and Scalogram with the highest scores of 71.39/16.66 in cough while still witness the second highest scores of 73.15/18.33 and 74.31/18.33 in speech and breathing, respectively.
As a result, we only use log-Mel spectrogram for evaluating network architectures.

As Table~\ref{table:pre_train_01} shows, it can seen that Xception outperforms the other network architectures. 
However, the high-complex Xception model only slightly improve, compared to the low footprint Lenet based architecture.
Notably, when high-level embedding features extracted from breathing, cough, and speech inputs by Xception model are concatenated to detect COVID-19 (i.e. The Track-4 in the Second DiCOVA Challenge), it  helps to improve significantly the performance, recording a score of 80.81/43.33 compared with 71.21/31.66, 70.91/28.33, 72.75/28.33 for only cough, speech, and breathing, respectively.

\begin{figure*}[th]
	\centering
	\centerline{\includegraphics[width=0.9\linewidth]{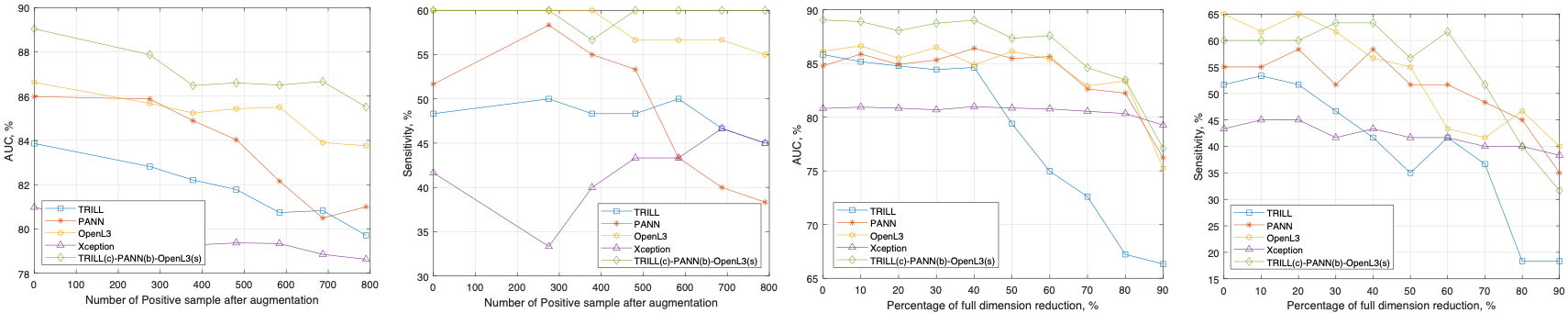}}
	 \vspace{0 cm}
	\caption{Influence of over-sampling on positive samples and dimensionality reduction using high-level features within Track-4 DiCOVA Challenge}
	\vspace{-0.3 cm}
	\label{fig:affect}
\end{figure*}	

\subsection{Performance comparison of frameworks (II): low-level log-Mel spectrogram, pre-trained models trained on AudioSet dataset, and LightGBM back-end classification}
\label{framework2}

As Table~\ref{table:pre_train_02} shows, it can be seen that different pre-trained models works well on different types of sound input.
Particularly, the best scores of 79.82/48.33 are associated with TRILL for cough sound input. PANN achieves the best scores of 84.05/55.00 for breathing. Meanwhile, speech sound input presents the best scores of 86.65/61.67 with OpenL3.
Notably, using pre-trained models with AudioSet dataset outperforms the frameworks (I) analysed in Section~\ref{framework1} in terms of detecting COVID-19 with only breathing, cough, or speech.

Comparing the performance of various models among audio inputs, breathing and speech show potential to detect COVID-19 rather than cough.
However, when we concatenate high-level embedding features extracted from breathing, cough, and speech for Track-4 of the challenge, the performance from TRILL shows a significant improvement to 85.83/51.67. Meanwhile, OpenL3 demonstrates the highest scores of 86.14/65.00 compared to the others in this track.
\begin{table}[t]
	\caption{Comparison of pre-trained models trained on AudioSet dataset (AUC/SEN.)} 
        	\vspace{-0.2cm}
    \centering
    \scalebox{0.85}{
    \begin{tabular}{|c  | c c c | c |} 
        \hline 
         \textbf{Pre-trained model}      &\textbf{Cough} &\textbf{Speech} &\textbf{Breathing} &\textbf{All} \\
        \hline 
	    PANN                 &70.49/26.66     &79.33/33.33  &\textbf{84.05/55.00}   &84.77/55.00 \\
	    TRILL                   &\textbf{79.82/48.33}     &80.50/46.67 &76.97/33.33 &85.83/51.67 \\
	    OpenL3                 &78.42/38.33     &\textbf{86.65/61.67} &82.40/50.00  & \textbf{86.14/65.00} \\
       \hline 

    \end{tabular}
    }
    \vspace{-0.3cm}
    \label{table:pre_train_02} 
\end{table}
%
\subsection{Influence of over-sampling on positive samples and dimensionality reduction of high-level features}

As the proposed frameworks with a concatenation of high-level features prove effective to detect COVID-19 in Track-4 of the Second DiCOVA Challenge,  we further conduct experiments to evaluate how over-sampling on positive samples and reducing dimension over high-level features affect the performance.
In these experiments, we concatenate (1) three TRILL based high-level features, (2) three PANN based high-level features, (3) three OpenL3 based high-level feature, (4) three Xception based high-level features, and (5) (TRILL(c)-PANN(b)-OpenL3(s)) high-level features for cough, breathing, and speech, respectively. 
We conduct the combination among high-level features (TRILL(c)-PANN(b)-OpenL3(s)) as each feature shows effective for each audio data input as shown in Table~\ref{table:pre_train_02}.

To identify less significant dimensions of high-level features, we firstly compute an average of each dimension from all features associated with positive and negative COVID-19 independently. Then, two vectors representing for two groups of COVID-19 positive and negative features are obtained. 
Based on the absolute difference of these two vectors from each dimension (i.e. the lower absolute difference is related to the less significant dimension), the dimension of the feature set has been reduced from 10\% to 90\%.  
Note that the dimension of high-level features is reduced before concatenating.
To over-sample the positive cases, we apply SVM-SMOTE~\cite{nguyen2011borderline} on the high-level embedding features, then increase the the positive samples from double to five times. 
 
As Figure~\ref{fig:affect} shows, oversampling on positive samples reduces the AUC scores. This method only helps to improve SEN. scores with PANN (double of positive samples) and Xception (five times of positive samples).
Meanwhile, when we reduce the dimension of high-level embedding features from 20 to 40\%, it almost helps to improve both AUC and SEN. scores. 
As a result, we can achieve the best scores: 80.32/48.33 with TRILL (at the reducing dimension of 20\%) for cough input only,  84.05/55.00 with PANN for breathing input only (no reduction), 86.65/61.67 with OpenL3 for speech only (no reduction), and 89.03/63.33 with (TRILL(c)-PANN(b)-OpenL3(s)) (at the dimensionality reduction of 40\%) for all audio inputs. 
\subsection{Performance comparison among different back-end classifiers}

We further compare the LightGBM with other machine learning models implemented by using Scikit-Learn toolkit~\cite{pedregosa2011scikit} on  of Support Vector Machine (SVM with C=1.0, Kernel=`RBF'), Random Forest (RF with Max Depth of Tree = 20, Number of Trees = 100), Multi-layer perceptron (MLP with 4096 nodes, Adam optimization, Max iter = 200, Learning rate = 0.001, Entropy Loss), and Linear Regression (LR with C=1.0).
Noting that we keep TRILL based embedding (at the dimensionality reduction of 20\%) for cough, PANN and OpenL3 based embeddings (with no reduction) for breathing and speech, respectively. Similarly, we continue to apply the best concatenation of (TRILL(c)-PANN(b)-OpenL3(s)) (at the dimensionality reduction of 40\%) for all audio inputs. As a result, the comparison in Table~\ref{table:Compare_ML_models} again shows that LightGBM model still achieves the best scores and outperforms other models.

\subsection{Performance comparison across the top-10 systems submitted for the Second DiCOVA Challenge}
\begin{table*}[t]
    \caption{Performance comparison with the state-of-the-art systems on Test set} 
        	\vspace{-0.2cm}
    \centering
    \scalebox{0.8}{
    \begin{tabular}{|c| c c c c|c c c c|c c c c|c c c c|} 
        \hline 
             &\multicolumn{4}{|c|}{\textbf{Cough}} &\multicolumn{4}{|c|}{\textbf{Speech}} &\multicolumn{4}{|c|}{\textbf{Breathing}} &\multicolumn{4}{|c|}{\textbf{All}} \\
        \hline 
             \textbf{System} &AUC  &SEN.  &Precision &F1 score  &AUC  &SEN.  &Precision &F1 score &AUC  &SEN.  &Precision &F1 score &AUC  &SEN.  &Precision &F1 score\\
             \hline
             1st system &81.97  &36.67  &52.38 &43.14  &85.21  &45.00  &57.45 &50.47 &87.18  &48.33  &59.18 &53.21 &88.44  &58.33  &63.64 &60.87\\
             2nd system &81.21  &48.33  &59.18 &53.21  &84.73  &38.33  &53.49 &44.66 &86.72  &40.00  &54.55 &46.16 &87.26  &63.33  &65.52 &64.41\\
             3rd system &80.12  &35.00  &51.22 &41.58  &84.55  &51.67  &60.78 &55.86 &86.41  &45.00  &57.45 &50.47 &86.87  &58.33  &63.64 &60.87\\
             4th system &79.06  &35.00  &51.22 &41.58  &84.26  &43.33  &56.52 &49.05 &85.77  &41.67  &55.56 &47.62 &85.79  &33.33  &50.00 &40.00\\
             5th system &77.85  &46.67  &58.33 &51.85  &84.04  &48.33  &59.18 &53.21 &84.50  &31.67  &48.72 &38.39 &85.37  &60.00  &64.29 &62.07\\
             6th system &77.60  &33.33  &50.00 &40.00  &82.98  &43.33  &56.52 &49.05 &82.16  &41.67  &55.56 &47.62 &84.70  &55.00  &62.26 &58.41\\
             7th system &76.98  &25.00  &42.86 &31.58  &82.83  &38.33  &53.49 &44.66 &82.05  &41.67  &55.56 &47.62 &84.26  &38.33  &53.49 &44.66\\
             8th system &76.36  &30.00  &47.37 &36.74  &81.95  &48.33  &59.18 &53.21 &80.84  &40.00  &54.55 &46.16 &83.78  &46.67  &58.33 &51.85\\
             9th system &75.95  &28.33  &45.95 &35.05  &81.86  &31.67  &48.72 &38.39 &80.55  &30.00  &47.37 &36.74 &80.51  &40.00  &54.55 &46.16\\
             10th system &75.71  &35.00  &51.22 &41.58  &80.92  &45.00  &57.45 &50.47 &80.35  &36.67  &52.38 &43.14 &74.15  &35.00  &51.22 &41.58\\
             \hline
             Baseline system &74.89  &36.67  &52.38 &43.14  &84.26  &43.33  &56.52 &49.05 &84.50  &31.67  &48.72 &38.39 &84.70  &55.00  &62.26 &58.41\\
             \hline
             \textbf{Our system} &\textbf{80.32}  &\textbf{48.33}  &\textbf{59.18} &\textbf{53.21}  &\textbf{86.65}  &\textbf{61.67}  &\textbf{64.91} &\textbf{63.25} &\textbf{84.05}  &\textbf{55.00}  &\textbf{62.26} &\textbf{58.41} &\textbf{89.03}  &\textbf{63.33}  &\textbf{65.52} &\textbf{64.41}\\
             
       \hline 
    \end{tabular}
    }
        \vspace{-0.3cm}
    \label{table:compSOTA} 
\end{table*}
\begin{table}[t]
	\caption{Performance comparison across different back-end classification models on Test set (AUC/SEN.)} 
        	\vspace{-0.2cm}
    \centering
    \scalebox{0.85}{
    \begin{tabular}{|c  | c c c | c |} 
        \hline 
         \textbf{Model}      &\textbf{Cough} &\textbf{Speech} &\textbf{Breathing} &\textbf{All} \\
        \hline 
	    LR                 &76.22/35.00     &80.90/40.00  &78.61/43.33   &77.29/43.33 \\
	    SVM                   &78.40/41.66     &85.93/53.33 &82.25/51.66 &85.73/55.00 \\
	    RF                 &78.78/43.33     &81.06/48.33 &79.06/38.33  & 85.34/51.66 \\
	    MLP                 &74.80/41.66     &81.53/56.66 &80.89/41.66  & 83.75/46.67 \\
	    LightGBM                 &\textbf{80.32/48.33}     &\textbf{86.65/61.67} &\textbf{84.05/55.00}  & \textbf{89.03/63.33} \\
       \hline 
    \end{tabular}
    }
        \vspace{-0.4cm}
    \label{table:Compare_ML_models} 
\end{table}
As performance comparison to the state-of-the-art systems~\cite{leaderboard} is shown in Table~\ref{table:compSOTA}, we can achieve the top-6 in Track-1 with breathing input only and the top-3 in Track-2 with cough input only.
Notably, our proposed systems outperform the state-of-the-art, achieve the top-1 in both Track-3 and Track-4 with speech input only and all audio inputs, respectively.

To evaluate the best score of the LightGBM with the best-selected embedding features for cough, speech, breathing and all types of audio inputs, we conducted ten times of running the experiments on a different randomly chosen Test set. Next, we calculate and achieve an average confidence interval (CI) of [0.7628, 0.8352] in cough, [0.8369, 0.8981] in speech, [0.8133, 0.8785] in breathing, and [0.8580, 0.9153] in all audio inputs. All of them matches with the mentioned AUC scores of 80.32, 86.65, 84.05, and 89.03 for cough, speech, breathing, and all, respectively.

\section{Conclusion}
\label{conclusion}
This paper has presented an exploration on how to extract effectively well-represented features for breathing, cough, speech sound input via pre-trained models. By conducting extensive experiments, we achieve a robust framework for detecting COVID-19 compared with the state-of-the-art systems. 
This is demonstrated by the rank of our performance with top-1 in both Track-3
and Track-4, top-3 in Track-2 and finally top-6 in Track-1 in the Second DiCOVA Challenge.
Our best AUC score of 89.03\%, F1 score of 64.41\%, sensitivity score of 66.33\% from the Track-4 demonstrate the potential of detecting COVID-19 through the respiratory-related sounds. 


\bibliographystyle{IEEEtran}
\bibliography{refs}

\end{document}